\documentclass[conference]{IEEEtran}
\pdfoutput=1
\usepackage{indentfirst}
\usepackage{amsmath}
\usepackage{mathtools}
\usepackage{amsfonts}
\usepackage{dsfont}
\usepackage{cite}
\usepackage{times}
\usepackage{graphicx}
\usepackage{subfigure}
\usepackage{bbm}
\usepackage{authblk}
\usepackage{enumerate}
\usepackage[noend]{algpseudocode}
\usepackage{xcolor}
\usepackage{booktabs}
\usepackage{caption}
\usepackage{authblk}
\usepackage{blindtext}
\usepackage{listings}
\usepackage{courier}
\usepackage[ruled, vlined, linesnumbered]{algorithm2e}

\newcommand{\figww}{0.7\columnwidth}

\title{eBPF-based Content and Computation-aware Communication for Real-time Edge Computing}
\author[1]{Sabur Baidya\thanks{sbaidya@uci.edu}}
\author[2]{Yan Chen\thanks{y.chen@huawei.com}}
\author[1]{Marco Levorato\thanks{levorato@uci.edu}}
\affil[1]{Donald Bren School of Information and Computer Science, UC Irvine}
\affil[  ] {e-mail: {\textit {\{sbaidya, levorato\}@uci.edu}}}
\affil[2]{America Software Laboratory, Huawei, e-mail: {\textit {y.chen@huawei.com}}}

\date{}

\makeatletter

\let\old@ps@headings\ps@headings
\let\old@ps@IEEEtitlepagestyle\ps@IEEEtitlepagestyle
\def\confheader#1{%
    \def\ps@IEEEtitlepagestyle{%
        \old@ps@IEEEtitlepagestyle%
        \def\@oddhead{\strut\hfill#1\hfill\strut}%
        \def\@evenhead{\strut\hfill#1\hfill\strut}%
    }%
    \ps@headings%
}
\makeatother

\confheader{%
        \parbox{\textwidth}{\centering This article has been accepted for publication in the IEEE International Conference on Computer Communications (INFOCOM Workshops), 2018.}
}

\usepackage[pscoord]{eso-pic}
\newcommand{\placetextbox}[3]{
\setbox0=\hbox{#3}
\AddToShipoutPictureFG{ \put(\LenToUnit{#1\paperwidth},\LenToUnit{#2\paperheight}){\vtop{{\null}\makebox[0pt][c]{#3}}}
}
}
\placetextbox{.5}{0.065}{\footnotesize {\copyright 2018 IEEE. Personal use of this material is permitted. Permission from IEEE must be obtained for all other uses, in any current or future media} }
\placetextbox{.5}{0.055}{\footnotesize { including reprinting/republishing this material for advertising or promotional purposes, creating new  collective works, for resale}} 
\placetextbox{.5}{0.045}{\footnotesize { or redistribution to servers or lists, or reuse of any copyrighted component of this work in other works.}
}

\begin{document}

\pagestyle{empty}
\thispagestyle{empty}

\maketitle

\begin{abstract}

By placing computation resources within a one-hop wireless topology, the recent edge computing paradigm is a key enabler of real-time Internet of Things (IoT) applications. In the context of IoT scenarios where the same information from a sensor is used by multiple applications at different locations, the data stream needs to be replicated. However, the transportation of parallel streams might not be feasible due to limitations in the capacity of the network transporting the data. To address this issue, a content and computation-aware communication control framework is proposed based on the Software Defined Network (SDN) paradigm. The framework supports multi-streaming using the extended Berkeley Packet Filter (eBPF), where the traffic flow and  packet replication for each specific computation process is controlled by a program running inside an in-kernel Virtual Machine (VM). The proposed framework is instantiated to address a case-study scenario where video streams from multiple cameras are transmitted to the edge processor for real-time analysis. Numerical results demonstrate the advantage of the proposed framework in terms of programmability, network bandwidth and system resource savings.


\end{abstract}

\section{Introduction}
\label{sec:intro}

In the Internet of Things (IoT), interconnected sensors collect and transmit data for analysis to remote servers~\cite{jin2012network}. However, the delivery of content-rich data may incur delay or loss due to the limited network resources. Indeed, network congestion may impair the ability of the system to support real-time services, such as video surveillance, traffic monitoring, smart transportation or the virtual reality (VR) and augmented reality (AR)~\cite{ali2017energy}. For this family of applications, the cloudlets~\cite{verbelen2012cloudlets} and edge computing~\cite{hu2015mobile} paradigms mitigate the issue of latency by placing computation-capable devices within a one-hop wireless topology. However, in many network scenarios of interest, the coexistence of these demanding data streams with other services over constrained wireless networks necessitates new technical solutions.

Recent frameworks based on Software Defined Networks (SDN)~\cite{nunes2014survey} have demonstrated the ability to improve network resource management using dynamic flow control and Network Function Virtualization (NFV)~\cite{hawilo2014nfv}. At the communication level, Software Defined Radios (SDR)~\cite{arslan2007cognitive} have been used to dynamically adapt the parameters of wireless transmissions. However, the main challenge of effectively utilizing the available bandwidth to support real-time applications producing large volume traffic stands.
To this aim, the notion of Quality of Computing (QoC)~\cite{baidya2017netselect} has been recently proposed to relax interference constraints on IoT data streams and facilitate their coexistence.

In this paper, we propose a computation-aware communication control framework for real-time IoT applications generating high-volume data traffic processed at the network edge. Driven by QoC requirements, the framework provides real-time user-controlled packet replication and forwarding inside the in-kernel Virtual Machines (VM) using an extended Berkeley Packet Filter (eBPF) \cite{begel1999bpf+}. The implementation uses the concepts of SDN and NFV to achieve highly programable and dynamic packet replication. Resource allocation is semantic and content-aware, and, in the considered case, informed by the structure of data encoding. Numerical results are provided from real-world experiments demonstrating the enhanced adaptability and efficiency of the proposed solution.


The rest of the paper is organized as follows. In section~\ref{sec:ecomp}, we describe the real-time edge computing scenario and formulate the problem. Section~\ref{sec:rel} discusses related work. In Section~\ref{sec:arch}, we present the architecture of the proposed framework and in section~\ref{sec:proto}, we describe the computation-aware communication protocol and provide the implementation details. Section~\ref{sec:eval} provides numerical results validating the proposed approach. Section~\ref{sec:concl} concludes the paper.


\section{Edge Computing for Real-time Applications}
\label{sec:ecomp}



\begin{figure}[!t]
	\centering
	\begin{minipage}[b]{0.70\textwidth}
	\includegraphics[width=\figww]{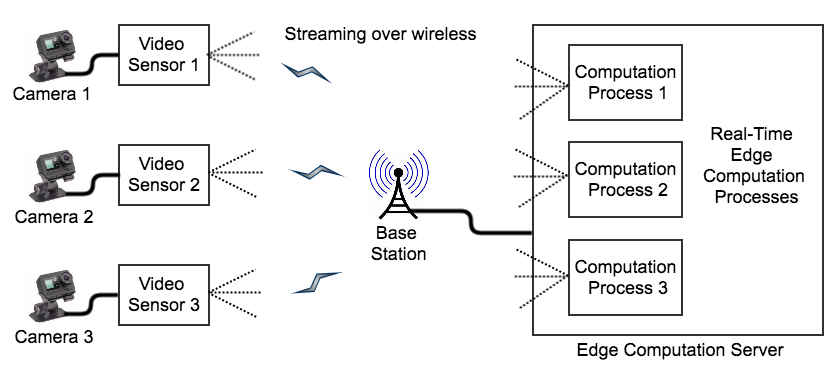}
	\end{minipage}
	  \vspace{-4mm}  
	\caption{Real-time edge-based scenario: each computation process uses partial data from various sensors.}
        \label{fig:usecase}
        \vspace{-1.5em}        
\end{figure}

One of the core concepts used in this paper is edge computing, which can
reduce the computation latency of real-time applications. It is generally assumed that the edge servers are more powerful compared to the sensors, but offer inferior performance compared to cloud servers in terms of computation capabilities. 
We consider an application scenario where video data from different video sensors (cameras) are sent to the edge server, which runs in parallel several real-time processes analyzing the streams as shown in Fig.~\ref{fig:usecase}.  Each computation process may use partial data from multiple sensors to accomplish its task. For instance, a computation process aimed at fine-grained object detection may require high quality video whereas coarse-grained object detection (e.g. object counting) can operate on relatively lower quality and lossy video. We remark that this setup corresponds, for instance, to Urban IoT scenarios and, in particular, city monitoring.

\vspace{2mm}

Thus, in the considered multi-sensor multi-computation scenario, each sensor data might be reused at different system scales for different computational tasks. Note that
if the sensors are mobile, the subset of sensors contributing to a specific computation process may also change over time. The ``quality'' of data a sensor should provide to a specific application is determined by the QoC requirements of the associated computation algorithm. Here, we define these requirements in terms of a set of metrics measuring characteristics of data delivery, such as loss and delay.



Let say, then, that at time $T$, $N$ video sensors $\{S_{1},S_{2},...,S_{N}\}$ are streaming video to $M$ computation processes $\{C_{1},C_{2},...,C_{M}\}$ running at the edge server. The $i_{th}$ sensor data stream $S_{i}$ is used by the $j_{th}$ computation process $C_{j}$ with quality $\Omega_{ij}$, where $i=1,2,...,N$ and $j=1,2,...,M$. In a naive approach, each sensor $S_{i}$ would transmit $M$ streams of data over the network with full transmission quality $Q_{i}^{full}$. Denote the bandwidth consumption stream $S_{i}$ with quality $Q$ as $S_i(Q)$ with $B_{i}$. The total bandwidth consumption for transmission with full quality is:
\begin{equation}
\vspace{-1mm} 
\small
B_{total} =  \sum_{i=0}^{N} \sum_{j=0}^{M}  {S_{i}(Q_{i}^{full})} 
{=} M \sum_{i=0}^{N}   {S_{i}(Q_{i}^{full})}       
\vspace{-1mm} 
\end{equation}

As an alternative approach, instead of sending $M$ instances of $S_{i}(Q_{i}^{full})$ to $M$ processes, the data streams are replicated in-network -- that is, at the network edge -- and sent to the $M$ processes in order to reduce the network load to {\small $B_{total} =  \sum_{i=0}^{N} {S_{i}(Q_{i}^{full})}$}. Thus, the number of streams is reduced to the minimum, but they are still transmitted at maximum quality $Q^{full}$ to support any requirement the computation processes may impose. In the framework we propose, we further reduce the network load by dynamically tuning the quality of the data streams produced by the sensors to that required by the computation processes at any given time.
 
In the framework proposed herein, packets are replicated inside an in-kernel VM at the data link layer (layer 2)  
before forwarding the packet to upper layers. This results in a decreased use of memory at the upper layers of the network protocol stack.
Our framework further reduces the system memory and CPU usage by incorporating content-based selective packet cloning, which is driven by the requirements of the individual computation processes.

In the light of these functionalities and objectives, in this paper we address the following technical challenges:
\begin{enumerate}[i)]
\item Determine the quality of the data streams transmitted by each sensor at run-time.
\item Select which of the data streams received at the network edge need to be forwarded, and to which computation process(es) at any point of time.
\item Efficiently manage network and computation resources used by each computation process.
\end{enumerate}
In order to solve the aforementioned challenges, we propose an approach based on SDN whose goal is to seamlessly control the resources of the network from the user space at the application layer.





\section{Related Work}
\label{sec:rel}
Several recent contributions focus on real-time IoT data streaming in edge computing architectures. A real-time video surveillance application is proposed in \cite{baidya2016content} where the transmission policy of the sensor is interference aware to facilitate coexistence with other communications. 
In~\cite{mao2017stochastic}, the authors adopt a stochastic optimization approach to maximize the efficiency of bandwidth usage by efficiently offloading computation to edge servers. However, the architectures presented in these works are not programmable. 
In~\cite{baidya2017content}, the authors present an edge-assisted SDN-based framework for the dynamic selection of the network used to transport data from real-time applications, and ~\cite{chen2016integrated} proposes an SDN-based resource management framework to perform content caching and server selection. However, the aforementioned works do not address content reusability and flow orchestration framework at the network edge. 

Some recent contributions proposed flow orchestration schemes using NFV at the network edge. Examples include NetFATE~\cite{lombardo2015open}, a framework based on Open vSwitch (OVS) \cite{pfaff2015design} running at the mobile edge. In~\cite{kulkarni2017nfvnice}, the authors proposed the platform NFVnice, which is built on OpenNetVM for the user space-controlled scheduling of NFV chains. However, while providing some level of flow control, these frameworks  do not provide sufficiently low level control to extract in real-time for the content transported by the data streams and inform content- and computation-based policies. Different from these contributions, the framework proposed herein adopts a content and computation-aware approach for resource management and dynamic control, and realizes an adaptable and programmable user-controlled platform. 
\section{Proposed Architecture}
\label{sec:arch}

To develop the proposed framework, we use a built-in kernel feature called extended Berkeley Packet Filter (eBPF).
In the following, we briefly introduce eBPF and, then, describe in detail the framework.

%

Packet filtering was first proposed by Mogul et al. \cite{mogul1987packer} to provide user-controlled filtering of a subset of packets in the network stack of kernel which is known as CMU/Stanford packet filter (CSPF). 
The Berkeley Packet Filter (BPF) \cite{mccanne1993bsd} has been developed to overcome the stack-based instructions and tree-based filtering model of CSPF. BPF employs a register based instruction set with directed acyclic control flow graph (CFG) for filtering mechanism which provides a significant performance gain over CSPF as CFG can retain the packet parsing states to avoid redundant comparisons. The BPF filter was implemented in Linux kernel 2.1.75 which supports 32 bit registers for instructions. However, a recent development of BPF called (eBPF) \cite{gregg2015ebpf} in kernel 3.18 and above, further improves its performance by introducing ten 64 bit registers and employing the JIT (just-in-time) compiler. Also, eBPF programs can be invoked from different layers of the network stack, e.g. socket, qdisc, and drivers. This latter feature enables BPF to process the captured packets before forwarding them to the subsequent layer, and load user defined program inside the BPF in-kernel virtual machine to process the traffic dynamically. Also eBPF maps can be shared between the user and kernel space to enable seamless control. The recent development of BPF within the context of the open-source project IO Visor \cite{oivisor2015} resulted in the BPF compiler collection (bcc) with LLVM back-end and clang front-end. The main advantage is that the front-end code can be written in high level languages such as python or C, which is translated in the bytecode by the BPF compilers to be loaded inside in-kernel virtual machine.

\begin{figure}[!t]
	\centering
	\hspace{-4mm}
	\begin{minipage}[b]{0.70\textwidth}
	\includegraphics[width=\figww]{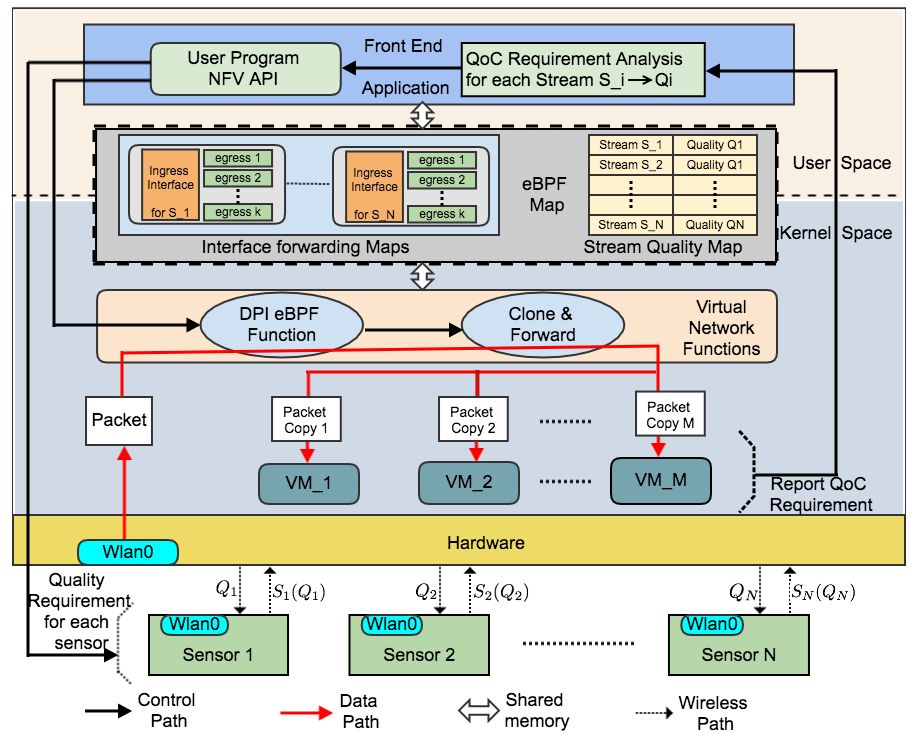}
	\end{minipage}
	\vspace{-4mm}
	\caption{Schematics of the content- and computation-aware stream control framework proposed in this paper.}
        \label{fig:clonearch}
        \vspace{-1.5em}        
\end{figure}

Based on these recent advancements, the schematics of the proposed framework are shown in Fig.~\ref{fig:clonearch}, whose components are described in the following.

\vspace{2mm}
\noindent
\emph{\underline{Quality of Computing Requirements}:}
The QoS requirements of each computation process are continuously evaluated at the communication and computation layers performance over a temporal window.  In the considered case, the QoC requirements are used to determine the minimum quality and portions of the data stream needed by the computation process. This parameter is reported by the computation processes to the application program, which updates the corresponding eBPF maps.
The edge, in turn, reports it back to the corresponding sensor, which only transmits the required portion of the data stream and suppresses remaining portions to reduce the network bandwidth used by the stream. 
Based on the determined requirements, the appropriate packets to be transmitted are identified by the sensor using a Deep Packet Inspection (DPI) module~\cite{baidya2017netselect} at the application layer. 

\vspace{2mm}
\noindent
\emph{\underline{Layer 2 forwarding}:}
Traffic forwarding to specific set of interfaces can be achieved by ingress/incoming and egress/outgoing port mapping through eBPF. In the proposed implementation, the user space application dynamically sets the ingress and egress port pairs in the eBPF map.


\vspace{2mm}
\noindent
\emph{\underline{DPI eBPF Function}:}
A DPI module is also implemented at the edge filters to select the packets to be forwarded to the computation process(es). This feature enables the implementation of content and computation-specific filtering policies. The DPI eBPF module runs inside the kernel, but can be accessed from the user space in real-time. 
 


\vspace{2mm}
\noindent
\emph{\underline{Selective Packet Cloning and Transmission}:}
The function of this module is to reduce the usage of computation resources at the edge servers. To this aim, the module employs content-aware selective cloning of packets to be forwarded toward a specific computation process, thus minimizing the socket buffer allocation of reusable packets. The eBPF program running inside the in-kernel VM reads the QoC requirement for each specific computation process through the eBPF maps, and makes decisions in real-time regarding the cloning policy and packet forwarding. 






\section{Content- and Computation-Aware Communication Control Protocol}
\label{sec:proto}

In this section, we make specific the optimization problem and the control strategy integrated in the framework to the application scenario we consider. 
We remind that the objective of the multi-sensor and edge system is to minimize total bandwidth usage, while meeting the QoC requirements, that is, maintaining the QoC metric $\gamma_{j}$ above a threshold $\delta_{j}$ for each computation process $C_{j}$. If each sensor uses a single stream $S_{i}$  with transmission quality $Q_{i}$ as described earlier, the optimization problem can be expressed as:
\begin{equation}
\label{optprob}
\min_{Q_{i}}  \sum_{i} \mathit{S_{i}(Q_{i})} ~~{\rm s.t.}~~  \mathit{\gamma_{j}}{\geq} {\delta_{j}}, ~\forall i=1,..,N;~\forall j=1,..,M.
\end{equation}

To solve this problem, we first make the communication between the sensor and edge computation-aware. The sensor determines a transmission quality $Q_{i}$ that satisfies $\Omega_{ij}$ for all $j=1,2,...,M$. Thus, we define effective quality of stream $S_{i}$ as:
\vspace{-1mm}
\begin{equation}
\vspace{-3mm}
\small
Q_{i}^{eff} = \bigcup\limits_{j=1}^{M} \Omega_{ij}
\end{equation}
\vspace{-1mm}
Intuitively, $Q_{i}^{eff}$ is smaller than or equal to $Q_{i}^{full}$. The reduction in terms of network usage can be expressed as the difference between $B_{total}$ and effective access network bandwidth $B_{eff}$:
\vspace{-2mm}
\begin{equation}
\vspace{-1mm}
\small
\begin{split}
\vspace{-3mm}
B_{saved}
= \sum_{i=0}^{N} {S_{i}(Q_{i}^{full})}~-~\sum_{i=0}^{N} {S_{i}( Q_{i}^{eff})} \\
\end{split}
\vspace{-2mm}
\end{equation}


We further reduce $Q_{i}^{eff}$ by making the communication content-aware. In the context of the specific application scenario considered in this paper, the video data stream is differentially encoded. As a result, the data stream is composed of packets transporting information relative to reference frames (full pictures) or differential frames (encoding differences with respect to reference frames).
In~\cite{baidya2017netselect}, we provided a preliminary study on the effect of packet loss on the performance of object detection algorithms. Our work illustrates the much larger impact of packet loss localized in portions of the stream transporting reference frames compared to the loss of packets transporting differential frames. The solution we proposed concentrates interference on differential frames, rather than equally spreading it over the data stream.

In this paper, we use this observation to build a content-aware packet filtering strategy. Specifically, when a reduction in the overall used bandwidth is necessary, the filter first drops packets that are transporting differential frames instead of uniformly dropping packets. Herein, we use a pre-built map, shown in Table 1, between the QoC requirement and packet loss in the two classes of content -- that is, reference and differential frames. The advantage in terms of used bandwidth to achieve a given object detection accuracy is apparent. Object detection is performed using the Speeded Up Robust Feature (SURF) detection~\cite{bay2006surf}. We summarize the steps of the filtering strategy and describe how content and computation-aware communications are implemented at the edge server and sensor side in Algorithm 1.





\begin{algorithm}
\small
\KwIn{ {\it ${\bf S} = \{S_{i} : i = 1,2,...,N\}$} : Sensors\\
\hspace{9mm} {\it ${\bf C} = \{C_{j} : j = 1,2,...,M\}$} : Computation processes\\
\hspace{9mm} ${\bf \Omega}[i,j]$ : Quality Requirement of $S_{i}$ for $C_{j}$}
\KwOut{Sensor to Computation Map : $\mathcal{G} : S \mapsto C$ \\
\hspace{10mm} ${\bf Q}[n]$ : Sensor Data Transmission Quality \\
\hspace{10mm} ${\bf \Delta}[i,j]$ : Edge Suppression Factor }
\SetKwBlock{Begin}{Function}{end Function}

\Begin($\text{EdgeControl} {~(} {\bf S, C, \Omega}  {)} :$)
{
$\textit{$\{\mathbb{I}_{i}\}$} : \text{Interface list for $S_{i}$}$\\
    \For{~ $i~=~1~to~N$}
    {
        $\textit{$\{\mathbb{I}_{i}\}$} \gets \text{{\bf null}}$\\
    	\For{~ $j~=~1~to~M$}
	{
	    \uIf{ {\textit{$C_{j}$}~includes~stream~$S_{i}$ }}
	    {
	    	$\textit{$\{\mathbb{I}_{i}\} \gets \{\mathbb{I}_{i}\}  +  {\it I_{j}}$}$
	    }
	    \Else
	    {
	    	continue
	    }
	}
	$\textit{$Q_{i}^{eff} = \bigcup\limits_{j=1}^{M} \Omega_{ij}$}$ ; $\textit{${\bf Q}[i] = Q_{i}^{eff} $}$\\
	
	\For{~ $k~=~1~to~\left\vert{\mathbb{I}_{i}}\right\vert$}
         {
             $ \textit{$A_{k}$} : \text{Application on $k_{th}$ VM} $\\
             $ \textit{$\delta_{i}$} : \text{Packet loss tolerance of $A_{k}$ for sensor data $S_{i}$}  $
             $\textit{${\bf \Delta}[i,k] = \delta_{i}$}$
             $\textit{$\mathcal{F}$} : \text{Frame type obtained by DPI of $S_{i}$ for $A_{k}$} $\\
             \uIf{ $\mathcal{F}$ $\in$ {\it Reference Frame}}
             {
                 $\text{Clone and forward packets to interface~${\it I_{k}}$}$
             }
             \Else 
             {
		$\text{Apply ${\bf \Delta}[i,k]$ suppression at interface~${\it I_{k}}$}$
                 $\text{Clone and forward packets to interface~${\it I_{k}}$}$
             }
         }
    }
    
}

\Begin($\text{SensorControl} {~(} {\bf Q}[n]  {)} :$)
{
    \For{~ $i~=~1~to~N$}
    {
        $ \textit{$\alpha_{i}$} : \text{Loss tolerance of $S_{i}$ for transmission quality ${\bf Q}[i]$}  $
        $ \textit{$P_{i}$} : \text{List of packets to be suppressed using DPI} $
        $\text{ $S_{i} \gets  S_{i} - P_{i}$ ; \text{Transmit} $S_{i}$ } $
    }
}

\caption{\small Content \&  Computation-aware Communication}\label{euclid}
\vspace{-2mm}
\end{algorithm}

To implement the DPI eBPF function, and the cloning and forwarding function, we use Virtual Network Functions (VNF) which can run inside in-kernel VM. The VNFs  can be invoked by the application program using APIs as shown in Listing~\ref{apis}. This allows the dynamic modification of the policies in the shared BPF maps.
We further optimize the DPI VNF implementation by setting the MTU size as multiple of transport stream (TS) packet size (188 bytes); so that the module does not scan every byte of the UDP payload, rather, jumps from one TS header of 4 byte size to the next TS header in the packet. This minimizes the number of instructions needed in the BPF. 
The implementation can integrate any future VNF accessible though a BPF API at the user program.

\vspace{-1mm}
\begin{lstlisting}[caption={APIs for accessing eBPF VNFs}, label={apis}]
// Access eBPF functions through API  
 bpf = BPF(src_file = "ebpf_vnfs.c", debug=0)
function_DPI = bpf.load_func("vnf_DPI", BPF.SCHED_CLS)
 ...
function_clone_forward = bpf.load_func("vnf_clone_forward", BPF.SCHED_CLS)
 ...
  
// Access and update eBPF shared Maps
pol_map = bpf.get_table("policymap")
pol_map[pol_map.Key(idx)] = pol_map.Leaf(arr, 0)
 ...
\end{lstlisting}


\begin{table}[b]
\vspace{-4mm}
\begin{tabular}{p{5cm}p{6cm}p{6cm}p{6cm}}
\hspace{4mm}
\small
\makeatletter
           \def\rulecolor#1#{\CT@arc{#1}}
           \def\CT@arc#1#2{%
           \ifdim\baselineskip=\z@\noalign\fi
           {\gdef\CT@arc@{\color#1{#2}}}}
           \let\CT@arc@\relax
          \rulecolor{gray!50}
        \makeatother
     \begin{tabular}{@{}lllll@{}}
        \toprule
        Object Detection (\%) & 0.5 & 1.0  & 2.0 & 5.0\\
        \midrule
        Uniform packet loss & 0.95 & 0.84 & 0.46 & 0.1\\
       Differential packet loss & 0.99 & 0.96 & 0.74 & 0.4\\
        \midrule
        {} & Packet & loss & (\%) &{}  \\
        \bottomrule
        \end{tabular}     
\end{tabular}

\vspace{2mm}
\hspace{6mm} Table 1: Object detection as a function of packet loss when the uniform and selective packet dropping policies are used.
\end{table}


\label{sec:stocmod}
\label{sec:perf}

\vspace{-4mm}
\section{Performance Evaluation}
\label{sec:eval}

In this section, we first describe the experimental setup, and then provide numerical results assessing the performance of the proposed framework.

\vspace{-2mm}
\subsection{Experimental Setup}
We implemented the eBPF program on a network edge server with 8 core CPUs. All the VMs run on QEMU hypervisor and we used the docker version 1.11.2 to define network containers. The implementation is based on Linux kernel 4.7.  As sensor data stream, we used a real-world video of $640$x$360$ pixels and $372$ frames encoded in H.264/AVC format at $30$ frames per second.  The video is converted to transport stream (TS) with the ffmpeg tool and transmitted from the sensor to the edge over UDP. 



\subsection{Results}

We evaluate the performance of the framework in terms of programability, layer 2 forwarding performance, bandwidth utilization as a function of the computation requirement, system resource usage and scalability.  
We tested the real-time redirection of traffic to the corresponding interfaces with respect to dynamic modifications in the mapping between the network source and destination ports, and the programmability demonstrated to be seamless.  

\vspace{2mm}
\noindent
\emph{\underline {Layer 2 forwarding performance}:}\
As discussed previously, the selective cloning and forwarding mechanism is implemented at Layer 2 using the DPI module at the edge server. The program is executed inside the in-kernel VM which runs with low level instructions. As a consequence, the flow does not incur significant delay traversing through the upper layers of the network protocol stack. Moreover, the layer 2 forwarding function in eBPF is connected to the {\it qdisc} of the kernel similarly to the Linux bridge. Hence, the eBPF switch achieves analogous layer 2 forwarding speeds as Linux bridge. We compare them by measuring their average throughput using the ``packetgen'' utility setting the packet size to $512$ bytes. We further compare the two cases using TCP and UDP. Both eBPF and Linux Bridge achieve  $\sim1$ Gbps throughput over UDP and $\sim2.5$ Gbps throughput over TCP (without TCP Segmentation Offload (TSO) and Generic Segmentation offload (GSO)). 



\begin{figure}[!t]
	\centering
	\begin{minipage}[b]{0.65\textwidth}
	\includegraphics[width=\figww]{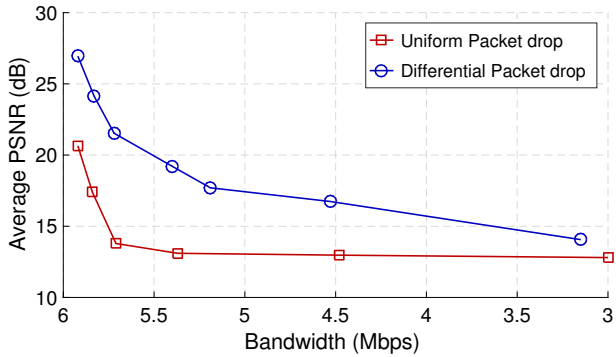}
	\end{minipage}
	\vspace{-5mm}
	\caption{PSNR as a function of channel capacity.}
        \label{fig:psnr}
        \vspace{-1.5em}        
\end{figure}

\vspace{2mm}
\noindent
\emph{\underline {Computation Performance}:}
We assess the performance of the proposed framework in terms of quality of the received video with respect to the network bandwidth utilization. First, we measure the average Peak Signal-to-Noise Ratio 
(PSNR) of the video by suppressing packets at the sensor. Fig.~\ref{fig:psnr} shows that as bandwidth usage is reduced by uniformly dropping an increasing number of packets, the PSNR decreases very sharply compared to the case where packet drop is focused on differential frames.
The DPI-based differential packet drop achieves significantly higher PSNR for any given bandwidth compared to the uniform, and non-selective, packet drop. For instance, when the available capacity of the communication link is equal to $5.5$~Mbps, the selective filtering strategy achieves a PSNR equal to $18$~dB, whereas the PSNR obtained using a uniform drop strategy is equal to $14$~dB. A PSNR equal to $15$~dB requires a capacity equal to $5.75$~Mbps and $3.6$~Mbps in the uniform and selective strategy, respectively.

\begin{figure}[!t]
	\centering
	\begin{minipage}[b]{0.65\textwidth}
	\includegraphics[width=\figww]{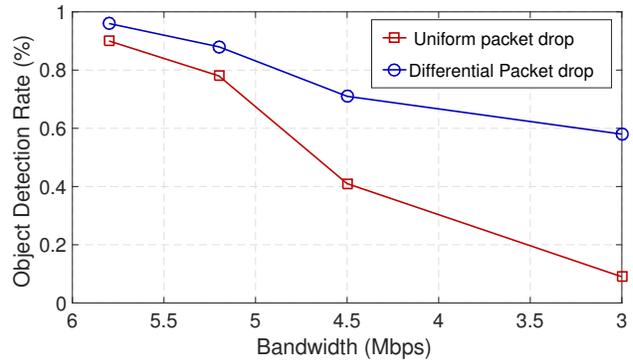}
	\end{minipage}
	\vspace{-5mm}
	\caption{Object detection rate as a function of channel capacity.}
        \label{fig:objdet}
        \vspace{-1.4em}        
\end{figure}

Further, we assess the performance of SURF-based object detection with respect to channel usage (see Fig.~\ref{fig:objdet}). The selective strategy provides a significant gain in terms of object detection rate compared to uniform packet drop for any given channel usage. For instance, when the available channel capacity is equal to $4.5$~Mbps, the object detection rate is equal to $0.4$ and $0.7$ in the uniform and selective drop strategy, respectively.


\begin{figure}[!t]
	\centering
	\vspace{1mm}
	\begin{minipage}[b]{0.65\textwidth}
	\includegraphics[width=\figww]{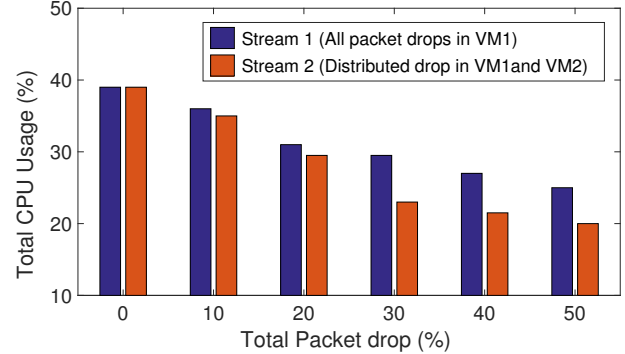}
	\end{minipage}
	\vspace{-5mm}
	\caption{CPU utilization as a function of packet drop rate.}
        \label{fig:cpuutil}
        \vspace{-1.6em}        
\end{figure}

\vspace{2mm}
\noindent
\emph{\underline {System Resource Utilization}:}
By selectively suppressing packets and avoiding cloning all the packets at the network edge, the proposed framework also improves system utilization. In the considered case-study, we use two video streams, each of which is processed by two processes running on two separate VMs (VM1 and VM2 respectively). We measure average CPU usage of all VMs with respect to packet drop rate. We distribute packet drop differently in the two streams: all the packets belonging to Stream 1 (first video) are dropped by VM1, whereas VM2 equally drops packets of Stream 1 and Stream 2.
Fig.~\ref{fig:cpuutil} shows that the CPU usage is reduced by $\sim 3\%$ on average for every $10\%$ total packet drop increase when all packet drops are performed on one VM. However, if packet drops are equally distributed between two VMs, then for every $10\%$ total packet drop increase, the CPU usage is reduced by $\sim 5\%$ on average. This indicates that the proposed framework provides an increasing advantage as the differences in QoS requirements of the computation processes increase.

\begin{figure}[!t]
	\centering
	\begin{minipage}[b]{0.65\textwidth}
	\includegraphics[width=\figww]{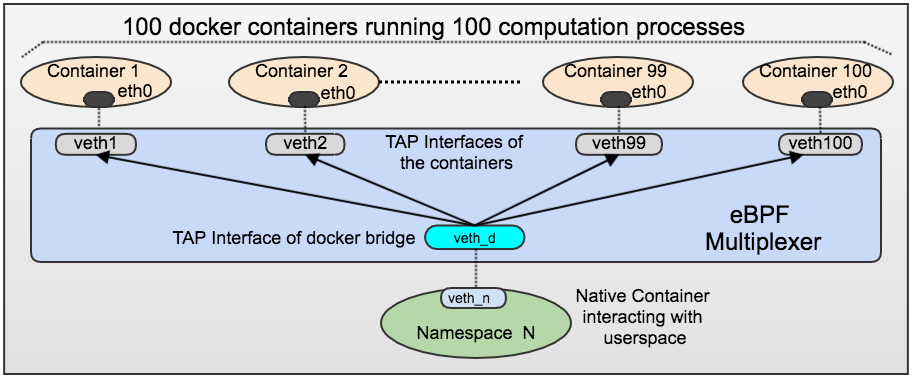}
	\end{minipage}
	\vspace{-4mm} 
	\caption{Multi-streaming test with 100 docker containers.}
        \label{fig:cont}
        \vspace{-1.0em}        
\end{figure}

\begin{figure}[!t]
	\centering
	\begin{minipage}[b]{0.65\textwidth}
	\includegraphics[width=\figww]{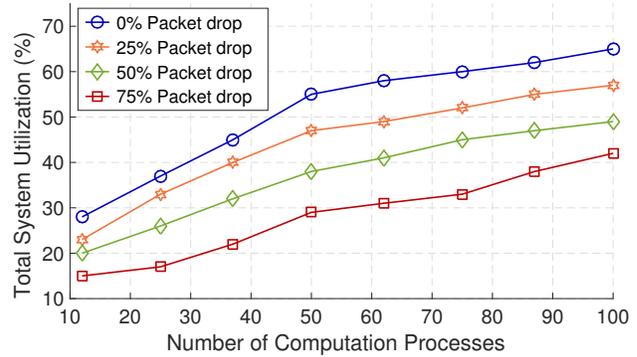}
	\end{minipage}
	\vspace{-4mm} 
	\caption{Variation of total system utilization vs number of containers for different packet drop percentage.}
        \label{fig:sysutil}
        \vspace{-1.8em} 
\end{figure}

\vspace{2mm}
\noindent
\emph{\underline{Scalability}:}
In order to test the scalability of the framework, we create a large number of docker containers for parallel computations (see Fig.~\ref{fig:cont}). On the sensor side, we use two sensor devices streaming data. The main goal of this test is to assess whether or not the framework can handle the execution of many parallel instructions in the kernel. Our tests show that due to the light-weight implementation of the DPI and selective clone-forward functions, and the low-level execution of our framework, the framework performs adequately even when the number of computation processes is large.
Fig.~\ref{fig:sysutil} shows that the total system utilization almost linearly increases with the number of containers. However, the slope decreases as the number of containers grows. This result indicates a tendency of the system resource utilization to become more efficient for large number of containers. Note that the total system utilization increases by $\sim 8\%$ for every $25\%$  increase in packet drop rate, that is, the efficiency increases as the load increases.

\vspace{-2mm}
\section{Conclusions}
\label{sec:concl}
The main contribution of this work is the design, implementation and test of an open-source, programmable computation-driven communication control framework for real-time edge computing systems using built-in kernel eBPF. The main features of the proposed system are: \emph{(a)} reduced network utilization to support computation processes; \emph{(b)} highly dynamic network and packet filtering control; and \emph{(c)} efficient system resource utilization at the edge server. Although the framework is demonstrated in a specific application, that is, video processing, the architecture is flexible and can support a broad range of IoT applications. Numerical results obtained by means of real-world testing demonstrate the ability of the proposed system to dynamically adapt data streams supporting remote computation processes.



\bibliographystyle{IEEEtran}
\bibliography{IEEEabrv,iot}

\end{document}